# Using Deep Space Climate Observatory Measurements to Study the Earth as An Exoplanet

Jonathan H. Jiang[1], Albert J. Zhai[2], Jay Herman[3], Chengxing Zhai[1], Renyu Hu[1], Hui Su[1], Vijay Natraj[1], Jiazheng Li[2], Feng Xu[1], Yuk L.Yung[1,2]

[1]Jet Propulsion Laboratory, California Institute of Technology, Pasadena, California
[2]Division of Geological and Planetary Sciences, California Institute of Technology, Pasadena, California
[3]Joint Center for Earth Systems Technology, University of Maryland, Baltimore County, Maryland

**Even though it was not designed as an exoplanetary research mission, the Deep Space Climate Observatory (DSCOVR) has been opportunistically used for a novel experiment, in which Earth serves as a proxy exoplanet. More than two years of DSCOVR Earth images were employed to produce time series of multi-wavelength, single-point light sources, in order to extract information on planetary rotation, cloud patterns, surface type, and orbit around the Sun. In what follows, we assume that these properties of the Earth are unknown, and instead attempt to derive them from first principles. These conclusions are then compared with known data about our planet. We also used the DSCOVR data to simulate phase angle changes, as well as the minimum data collection rate needed to determine the rotation period of an exoplanet. This innovative method of using the time evolution of a multi-wavelength, reflected single-point light source, can be deployed for retrieving a range of intrinsic properties of an exoplanet around a distant star.**

1. **Introduction**

An extrasolar planet, or exoplanet, is a planet outside our solar system, circling a star other than our sun. The first scientific detection of an exoplanet was in 1988 (Campbell et al. 1988). Since then, and as of May 8, 2018, there have been 3,725 exoplanets, in 613 multi-planetary systems, detected and confirmed according to the NASA Exoplanet Archive (https://exoplanetarchive.ipac.caltech.edu/). In addition, NASA's Kepler mission has identified nearly 4,500 extrasolar planetary candidates (Chou et al. 2017), several of them being nearly Earth-sized and located in the habitable zone, some around Sun-like stars. Atmospheres have also been detected around several exoplanets (Charbonneau et al. 2002), as well as the existence of an exomoon (Bennett et al. 2014).

So far, most of the exoplanets have been detected via indirect methods, such as measuring transits or starlight wobbles. A few of these exoplanets that have been directly imaged are mostly massive and are far from the glare of their star (Kalas et al. 2008; Currie et al. 2012), Among them, the least massive planet is Fomalhaut b, also known as Dagon, which has a mass less than twice Jupiter's mass and is orbiting the A-type main-sequence star Fomalhaut, about 25 light-years away in the Piscis Austrinus constellation (Currie et al. 2012).

In recent years, a few habitable or Earth-like exoplanets have been discovered, including some that may be orbiting Sun-like stars (Petigura et al. 2013; Gilster and LePage 2015). A habitable planet is a terrestrial planet within the circumstellar habitable zone, and with conditions roughly comparable to those of Earth so as to potentially favor water-based Earth-like life (Lammer et al. 2009). One of the most recent discoveries of such an Earth-like planet by NASA's Spitzer Space Telescope is TRAPPIST-1e, which is one of seven new exoplanets discovered orbiting the star TRAPPIST-1 approximately 40 light-years from Earth (Gillon et al. 2017). TRAPPIST-1e is located within the star's habitable zone and has an estimated equilibrium temperature of 251 K, which is very close to Earth's equilibrium temperature of 255 K. It has a radius that is 8% smaller than that of Earth radius and it is 38% less massive than Earth.

However, there is a significant gap between knowing the geophysical, geodynamical, and surface features of a habitable planet, and establishing whether it could harbor life. It requires direct imaging, spectroscopy, and development of techniques to extract fundamental planet properties, such as rotation, atmospheric composition, clouds, seasonal changes, surface features, and ultimately, biosignatures. Over the past decade, a NASA Terrestrial Planet Finder (TPF) mission has been proposed to detect light reflected by planets orbiting stars to investigate whether they could harbor life. The European Space Agency (ESA) had a similar mission proposal, named Darwin, to study the light from planets to detect the signatures of life. Both proposals would use advanced telescope technologies to look for biosignatures in the light reflected from planets. In preparation for these missions, astronomers have performed various Earthshine observations, using spectroscopic characteristics of reflected earthlight from the Moon's night side to emulate exoplanet observations (Sterzik et al. 2012; Arnold et al. 2002). The main challenge for the TPF and Darwin missions, however, is not gathering sufficient photons from the faint exoplanet for performing a spectral analysis, but rather detecting a faint point-light (exoplanet) that is close to a very bright star. In such a case, an alternative methodology is to detect signatures from the time evolution of a multi-wavelength reflected point-light source, which can be traced back to various intrinsic properties of the exoplanet. Studies of Earth images provide a demonstration of this method.

The reflected light of the integrated Earth disk, or "point-light", has been empirically measured from the ground via Earthshine (i.e., sunlight that has been reflected from Earth onto the dark side of the Moon and back again to Earth) and from a number of spacecraft. Earthshine experiments have detected strong diurnal variations in Earth's disk-averaged albedo (Goode et al. 2001; Pallé et al. 2003; Langford et al. 2009), and have measured Earth's typical optical to infrared spectra, loaded with molecular absorption features (Arnold et al. 2002; Woof et al. 2002; Turnbull et al. 2006; Sterzik et al. 2012). From space, the Galileo spacecraft heading to Jupiter and the Lunar Crater Observation and Sensing Satellite (LCROSS) observed Earth in a few snapshots, producing the ground-truth spectrum of Earth from visible to mid-infrared as viewed from afar (Sagan et al. 1993; Robinson et al. 2014).

NASA's EPOXI* mission measured Earth's reflected light and its diurnal variation. EPOXI observed Earth in three epochs each lasting for 24 hours, obtaining spectrophotometry in seven 100-nm wavelength bands from 300 to 1000 nm and spectroscopy from 1000 to 4500 nm (Livengood et al. 2011). The obtained dataset showed that the Earth spectra contain information about continents, oceans, and clouds (Cowan et al. 2009; Cowan et al. 2011), and that the infrared absorption features of $H_2O$, $CO_2$, and $O_2$ in the spectra vary due to uneven and changing cloud coverage (Fujii et al. 2013).

Numerical models have also been developed to predict Earth's reflected light spectrum and its variability (Ford et al. 2001; Tinetti et al. 2006). Fueled by the EPOXI observations, tremendous progress has been made on the inverse problem, i.e., inferring the surface geography of a planet based on time-resolved photometry. This includes the development of a principal component analysis (PCA) to decompose the EPOXI spectra into two dominant "eigencolors," red and blue, and an inversion of the diurnal light curves to derive a longitudinal map of these two surface components (Cowan et al. 2009; Cowan et al. 2011). The resulting map clearly shows the Atlantic and Pacific Oceans and major landmasses in between. Note that this technique does not assume any prior knowledge of the surface types or their albedo spectra. This analysis is expanded to include a third component attributed to clouds (Cowan and Strait 2013), but the degeneracy between the surface colors and their spatial distributions is found to be severe (Fujii et al. 2017). When assuming a template reflectance spectra of the major surface types including ocean, snow, soil, vegetation, and clouds, the EPOXI spectrophotometry can be decomposed to recover the fractions and longitudinal distributions of the various surfaces (Fujii et al. 2010; Fujii et al. 2011).

Despite this progress, the use of Earth images for exoplanet studies has been limited in scope. It has been suggested repeatedly using theoretical models, that Earth's rotational period can be estimated from its reflected light variations, even in the presence of time-varying clouds (Pallé et al. 2008; Oakley et al. 2009). Empirically testing this idea requires observations of a long temporal baseline. Furthermore, for an observational baseline that covers the orbital revolution of the planet, it has also been shown that the combined rotational and orbital variations of the planet's color spectrum can be used to derive two-dimensional maps of the planet's surfaces (Kawahara and Fujii 2010; Kawahara and Fujii 2011; Fujii and Kawahara 2012; Cowan & Fujii 2017; Farr et al. 2018; Haggard & Cowan 2018). The planet's obliquity may be simultaneously derived within this procedure, or even independently from solving the albedo map of the planet (Kawahara 2016).

Here we report the Deep Space Climate Observatory (DSCOVR) observations that span more than two years. We average Earth-observing DSCOVR's 10-wavelength reflected light signal into an equivalent single-point time series to emulate exoplanet signals. This study is the first empirical proof that intrinsic properties of an Earth-like exoplanet, including planet's rotation period, surface and cloud variations can be determined from reflected light spectrum variations.

2. **The DSCOVR Earth Images**

DSCOVR was launched on February 11, 2015, and has been positioned at the Sun-Earth $1^{st}$ Lagrangian ($L_1$) point (Lagrange et al. 1811), 1.5 million kilometres from the Earth, between the Sun and Earth, since June 2015. DSCOVR's Earth Polychromatic Imaging Camera (EPIC, https://epic.gsfc.nasa.gov) peers back at Earth and images the entire planet to detect changes in the planet's albedo, ozone absorption and clouds (Herman et al. 2017). DSCOVR is operated by the National Oceanic and Atmospheric Administration (NOAA) primarily to monitor space weather (e.g. solar storms). It is also equipped with two NASA Earth-viewing instruments: the National Institute of Standards and Technology Advanced Radiometer (NISTAR) and EPIC (described above). NISTAR is designed to measure the reflected and emitted energy from the entire sunlit face of our planet in three broad wavelength bands as a single pixel. EPIC provides high spatial resolution (18×18 km$^2$) spectral images of the Earth (10 wavelength bands, see Figure 1). Most recently in May 2017, glints of light from Earth, seen as twinkling in the DSCOVR's EPIC images, were confirmed to be reflected light from cloud ice crystals in the atmosphere (Marshak et al. 2017), in addition to specular reflections from oceans.

Although DSCOVR was not designed for exoplanetary study, its Earth images can be used to synthesize brightness signals of a rotating unresolved Earth observed from afar, thus simulating observational data of an exoplanet. Here we study the time series of the brightness data derived from DSCOVR EPIC's multiple channel images and use Fourier analysis techniques to extract periodic behavior due to planetary rotation, cloud variations, surface type (ocean, land, vegetation), and seasonal and annual changes. We also show that cloud and surface signals may be separated, since clouds have a relatively uniform response across the spectrum, while land features, especially vegetation, have varying albedo at different wavelengths measured by EPIC. The results from this study will greatly assist exoplanet research in the near future.

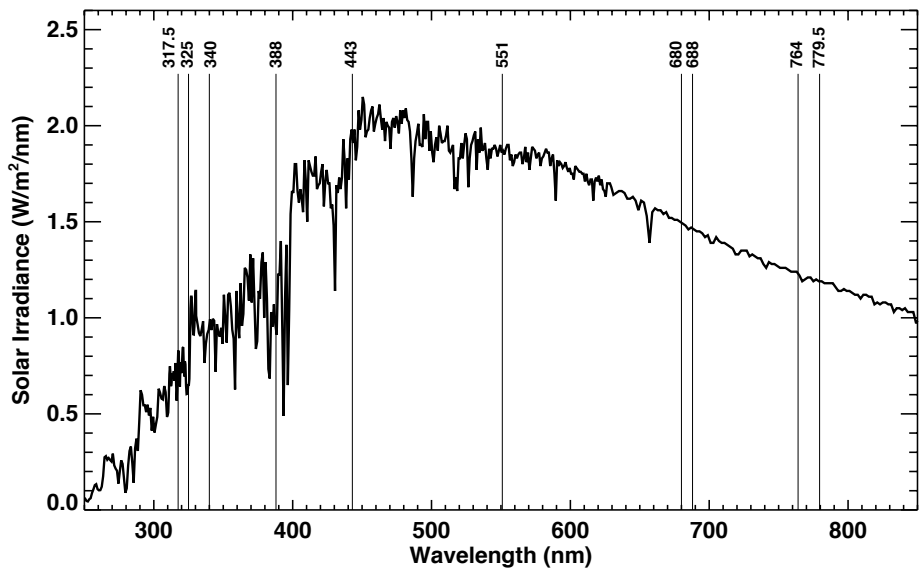

**Figure 1.** Solar irradiance spectrum and center wavelength positions for the 10 DSCOVR EPIC narrowband channels. The solar irradiance data are taken from the World Climate Research Program (WCRP).

The EPIC instrument consists of a 2048×2048 hafnium-coated Charge-Coupled Device (CCD) camera with 12-bit readout electronic (Herman et al., 2017). The images are taken with 10 narrowband filters, four in the ultraviolet (317.5, 325, 340, and 388 nm), four in the visible (443, 551, 680, and 688 nm), and two at the near infrared (764 and 779.5 nm) wavelengths. Two of these channels (688 and 764 nm) are within the strongly absorbing oxygen B- and A-bands. The filter widths, transmissions and quantum efficiency of the 10 spectral bands are shown in Table 1 (Herman et al. 2017; Geogdzhayev and Marshak 2017). The electronic light signals from the CCD are read out from the 12-bit analog-to-digital (A/D) converter with a gain of 42 electrons/count. EPIC's Level 1B (L1B) data are provided in engineering units of counts divided by the exposure time (counts/second) (Herman et al. 2017).

To convert the EPIC L1B images from units of counts/second to radiance units of $Wm^{-2}nm^{-1}$, we first use the EPIC calibration factors $K$ (Table 1) to convert counts/second ($ct$) into reflectance ($R$), according to:

$$R = ct \cdot K \qquad (1)$$

and then approximately convert to radiances ($I$), according to:

$$I = R \cdot E/\pi = ct \cdot K \cdot E/\pi \qquad (2)$$

where E is the solar irradiance in $Wm^{-2}nm^{-1}$ (Table 1 and Figure 1). The calibration factors $K$ are obtained using in-flight scene matching calibration from Earth orbiting satellites for most of the EPIC channels, and from the Moon for the two oxygen absorbing channels (688 and 764 nm) (Geogdzhayev et al. 2017).

**Table 1.** Instrument Parameters, Calibration Factors, and Solar Irradiance of EPIC Wavelength Band

| Wavelength (nm) | Filter Width* (nm) | Exposure (ms) | Transmission (%) | CCD Quantum efficiency (%) | Calibration Factor K (nm) | Solar Irradiance E (W/m²/nm) |
|---|---|---|---|---|---|---|
| **317.5** | 1.0 | 654 | 82.5 | 83.1 | 1.216×10⁻⁴ | 0.810 |
| **325** | 1.0 | 442 | 80.8 | 84.1 | 1.111×10⁻⁴ | 0.651 |
| **340** | 2.7 | 67 | 78.5 | 84.3 | 1.975×10⁻⁵ | 0.965 |
| **388** | 2.6 | 87 | 74.0 | 82.7 | 2.685×10⁻⁵ | 0.939 |
| **443** | 2.6 | 28 | 80.1 | 79.6 | 8.340×10⁻⁶ | 1.945 |
| **551** | 3.0 | 22 | 80.3 | 77.2 | 6.660×10⁻⁶ | 1.865 |
| **680** | 1.6 | 33 | 77.1 | 70.3 | 9.300×10⁻⁶ | 1.495 |
| **688** | 0.84 | 75 | 77.0 | 69.9 | 2.020×10⁻⁵ | 1.465 |
| **764** | 1.0 | 101 | 71.5 | 60.6 | 2.360×10⁻⁵ | 1.230 |
| **779.5** | 1.8 | 49 | 70.3 | 57.1 | 1.435×10⁻⁵ | 1.190 |

* The Filter Width is computed as Full Width at Half Maximum (FWHM).

EPIC has been taking Earth images since June 13, 2015 at a rate of one set of 10 wavelengths every 68 to 110 minutes, with exposure times ranging from 22 ms (551 nm) to 654 ms (317.5 nm). The field of view (FOV) is 0.61° with an angular sampling resolution of 1.07 arcsecs, and the camera produces 2048×2048 pixel images that are downsized onboard DSCOVR to 1024×1024 images except for the 443 nm channel. The downsized channels have a nadir resolution of ~18 km/pixel based on the instrument's point-spread function. The 443 nm channel has been shown to have a resolution of ~10 km/pixel at the image nadir point. An example of Earth imagery measured by the 10 EPIC wavelength channels at ~9:22 Coordinated Universal Time (UTC) on February 8, 2017 is shown in Figure 2. The images are from the EPIC L1B version-2 data (https://eosweb.larc.nasa.gov/project/dscovr/dscovr_table/), which have been converted into radiances (Wm$^{-2}$sr$^{-1}$) using Equation (2).

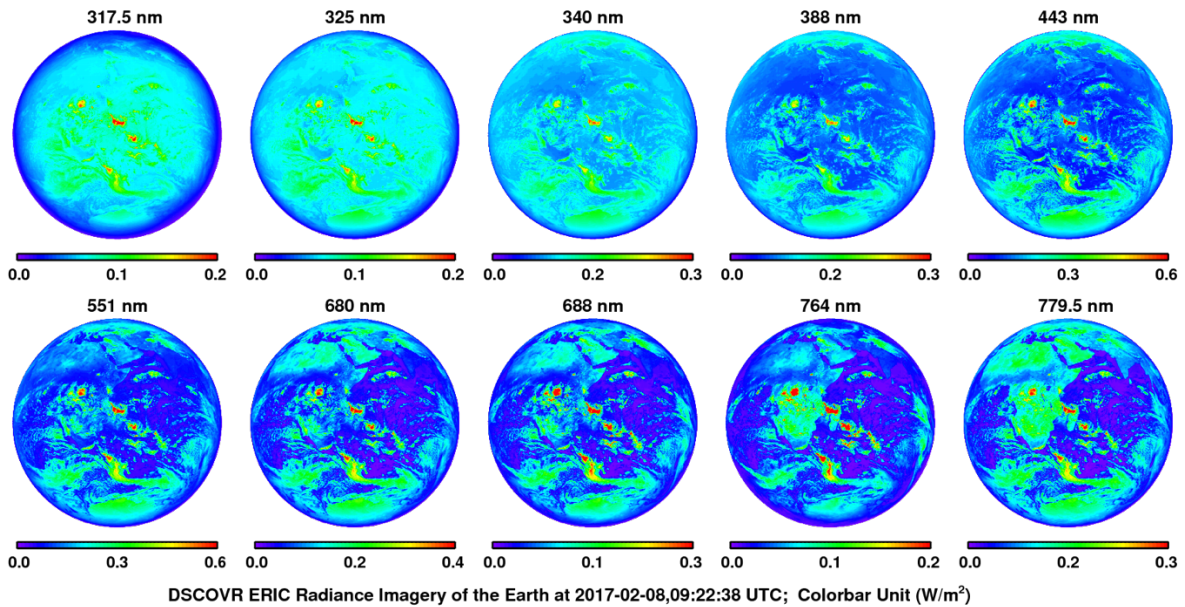

**Figure 2.** DSCOVR EPIC's 10 wavelength imagery of the Earth taken at 9:22 UTC, February 8, 2017.

There are a few interesting features in these reflected sunlit images of the Earth at different wavelength bands. First, the Earth looks brightest in the 443 nm and 551 nm EPIC images (note the images are displayed with different scales to improve visualization). This is because with Sun's surface temperature of 5778 K, the peak irradiance of sunlight is within the 450-550 nm wavelength range (Figure 1). Second, reflected light from clouds dominates all 10 wavelengths, whereas continents are most easily visible in the red and near infrared bands because of reduced Rayleigh scattering, especially at 780 nm wavelength where vegetation reflects strongly and the oceans are dark. Third, Rayleigh scattering plays a major role in contributing radiance in the UV bands over both land and ocean, because of the intensification of Rayleigh scattering with decreasing wavelength. Reflection from the oceans is significant (10-20%) in the blue and green bands, but decreases sharply in the UV and red and near-infrared bands.

Various substances reflect the Sun's radiation differently, causing the measured reflectance to vary with wavelength. The reflectance properties of an object depend on the material's physical and chemical state, as well as the surface roughness (Coakley 2003; Henderson-Sellers and Wilson 1983). Clouds, ice and snow generally have high reflectance across all of the 10 wavelength bands, hence their bright white appearance. These spectrally uniform features can easily be distinguished in the EPIC images: the distribution of clouds, especially thick clouds, is seen in all 10 wavelength bands. The ice cap of Antarctica is also clearly visible at all 10 wavelengths in Figure 2 (e.g., February or northern winter), when the southern hemisphere is tilted more toward the EPIC camera. In contrast, vegetation such as plants and forests is a strong absorber of electromagnetic energy in the UV, and a moderate absorber in the visible region. However, it reflects strongly in the near infrared between 700 nm and 1.3 μm, primarily due to the internal structure of plant leaves (Jacquemoud and Baret 1990). Sunlight reflected from soil also has a wavelength dependence. Bare soil generally has greater reflectance in the near infrared. Some of the factors affecting soil reflectance include moisture content, composition (e.g. sand, silt, clay, etc.), and surface roughness (Cierniewski and Verbrugghe 1997). This wavelength-dependent reflectance may be used to discriminate vegetation and soil from cloud features. For example, in Figure 2, features over the African continent are more easily seen in the 764 and 780 nm wavelengths.

Ocean, on the other hand, has greatest reflectance in the 388nm and blue-green wavelengths. The 318, 325, and 340 nm channels appear bright due to atmospheric Rayleigh scattering. Liquid water has high absorption in both the ultraviolet and red wavelengths, having a minimum absorbance in the blue, and virtually no internal reflectance in the near infrared and longer wavelengths (Pope and Fry 1997), other than surface Fresnel reflection.

These differences in reflectance between short and long wavelength channels discussed above, raise the possibility of identifying different Earth-like features in exoplanets by analyzing their spectral reflectance signatures.

### 3. Separating Clouds from Surface Features

To isolate surface features given the dominance of cloud reflected light in the data, we wish to remove the cloud irradiances by linearly combining two wavelength channels. This is expected to be possible because cloud reflectance is approximately wavelength independent for all types of clouds, while different surface types have varying reflectances in different spectral channels. For example, comparing 388 nm images with 780 nm images in Figure 2, it is clear that the highly reflective clouds dominate the 388 nm images against a relatively dark surface background. Therefore, the reflectance seen at 388 nm is a good representation of the cloud population. On the other hand, at 780 nm, contributions from surface features and clouds are both significant. It is then possible to remove the cloud irradiance in the image at 780nm ($I_{780}$) by proportionately subtracting the image at 388 nm ($I_{388}$).

To subtract clouds effectively, we look for the best factor $f$ so that the resulting signal, which we call the surface-signature radiance ($I_s$),

$$I_s = I_{780} - f \cdot I_{388} \qquad (3)$$

has a minimal cloud signal. In other words, the cloud irradiance signals at 388nm multiplied by the factor $f$ should closely match the cloud irradiance signals at 780nm. The factor $f$ is estimated using a regression over regions where the signals in the images come exclusively from clouds. To determine the regions where the irradiances are solely due to clouds, we choose a cutoff for the radiances above which only cloud pixels remain, because only clouds can reflect light at the high end of the radiance histogram. After experimenting with various percentile cutoffs, we chose the cutoff of the distribution of $I_{780}$ values to be at the 75th percentile, since it maximized the number of included cloud regions, while simultaneously excluding surface contributions. After all areas below this cutoff were masked out, the remaining $I_{780}$ values above the 75th percentile mainly contain clouds, which we call the *cloud-signature radiance* ($I_c$) (Figure 3, left-panel). A linear regression is then performed over the $I_c$ to calculate the value of $f$ that maximizes the removal of cloud contribution.

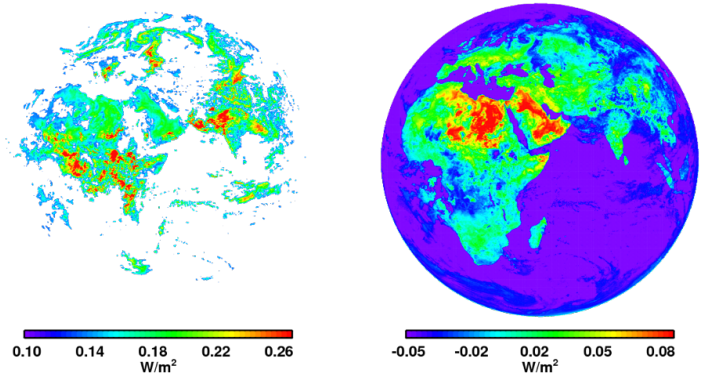

**Figure 3. Left:** Cloud signature radiance ($I_c$), which is the $I_{780}$ values above the 75[th] percentile from the EPIC image of 8:57 UTC, August 8, 2016. The 75[th] percentile of pixel measurements is a good approximation for a cutoff that maximizes cloud coverage while excluding all surface contributions. Using this cutoff, ~90% of cloud can be removed. **Right:** Surface signature radiance ($I_s$), computed using equation (3).

The mean of the scaling factors for all images is $f = 1.22$, which is used to scale the whole dataset to obtain an $I_s$ time series. The cloud areas' root mean square (RMS) after the subtraction divided by the same locations' RMS of cloud irradiance before the subtraction is used as a metric for how much cloud remained in the $I_s$. We estimate that less than 9% cloud reflectances remain by using the above cutoff method. Note the $I_s$ includes both landmass and ocean signatures: positive values over land and negative values in the ocean (Figure 3, right-panel).

This analysis demonstrates that the cloud contribution can be removed by linear combination of the spectral channels, with the same coefficients applied to every pixel of the entire globe. As such, we provide a pixel-level proof that the PCA and the spectral decomposition methods previously developed to analyze the disk-integrated time series of EPOXI (Cowan et al. 2009; Fujii et al. 2011) are sound. In the following we adopt this cloud-removal scheme to separate the components of land and cloud in the disk-integrated time series of DSCOVR/EPIC.

## 4. The "Single-Point Observation" Time Series

Until recently, there was no exoplanet optical imaging mission concept that could resolve earth-like exoplanets. The angular size $\theta$, of an exoplanet with radius $R$ at a distance $d$ is

$$\theta = 2 \cdot \arctan(R/d) \approx 2R/d. \qquad (4)$$

Imagine placing the Earth at the distance of the closest neighboring star, Alpha Centauri A, which is ~4.3 light years away. The angular size of the Earth would be ~$6.3 \times 10^{-5}$ arc-second, using the Earth radius of ~6400 km. In the visible regime, the wavelength is $\lambda$ ~0.6 μm, thus according to Rayleigh's resolution criteria 1.22 $\lambda/D$, we would need a telescope, or telescope array, of effective diameter $D$ ~2.3 km to resolve this, which is currently not available.

EPIC images from the DSCOVR mission provide a unique opportunity to monitor the Earth for an extended period in great detail. Averaging all of the EPIC data in each sunlit disk image into a single value allows us to emulate the signals from a distant exoplanet. Analyzing the variations of such point light sources in terms of the processes contained in the high resolution images provides insights in how to interpret the variations in a distant exoplanet point source for possible atmospheric and surface changes.

For each EPIC image, all data pixels that represented the Earth (and the Moon when it entered the FOV) were collected, and an arithmetic average was taken to simulate a single-point measurement. The time series of the mean single-point radiances for two single days, February 8, 2017 and August 8, 2016, are shown in the left column of Figure 4, and the time series of the entire EPIC data are shown in the right column of the same figure.

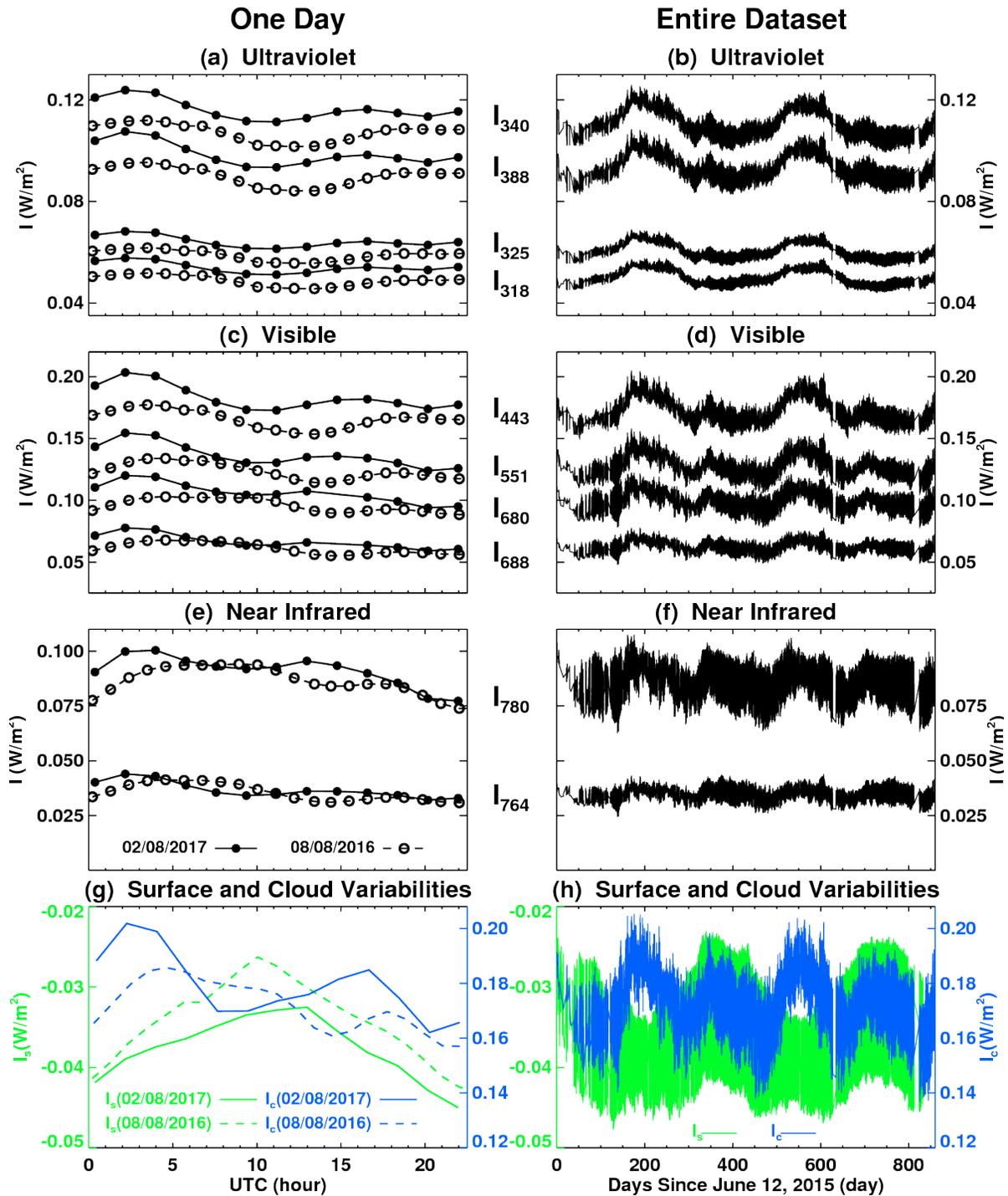

**Figure 4.** Time variations of DSCOVR EPIC L1B radiances after averaging to single-point measurements for each wavelength band, as well as surface and cloud signature radiances ($I_s$ and $I_c$). **Left:** daily measurements from February 8, 2017 (solid-lines and dots) and August 8, 2016 (dashed-lines and open circles); **Right:** entire time series of the single-point measurements over the full dataset from June 12, 2015 to October 17, 2017.

In the ultraviolet wavelength channels (Figures 4a and 4b), clouds contribute the majority of the reflected light. Whenever a cluster of clouds is in the EPIC FOV, the reflected radiance is enhanced. For example, between 2:00-4:00 UTC on both February 8 and August 8, when the Pacific Ocean is facing the EPIC camera, the large convective clouds near the maritime continents and the Inter Tropical Convergence Zone (ITCZ) across the Pacific, cause strong enhancements of the observed reflectance (Figure 4a). A secondary maximum occurs near 15:00-17:00 UTC when convective clouds over South America and Africa, and across the Atlantic Ocean are in the FOV (See Figure A1 in the Appendix). Oceans contribute only a small potion to the reflections except in the relatively small "glint" or specular reflection region. Increasingly strong Rayleigh scattering at 318 and 325 nm reduces the contrast from light reflected by clouds. In addition, the 318 and 325 nm reflected radiances are reduced because they are within the strong ozone absorption bands.

In the visible wavelength channels (Figures 4c and 4d), clouds still dominate the brightness of the Earth in a similar way, especially at 443 and 551 nm. In the two red wavelength channels, especially at 680 nm, the land contribution starts to mix with the cloud signals, thereby weakening the variability in the time series. The land contribution becomes more dominant *in the near infrared channels*, where the variability in the time series becomes very small (Figure 4e and 4f). Note that 688 and 764 nm channels are located in the $O_2$ B- and A-band absorption regions. The 764 nm solar irradiance is strongly absorbed by $O_2$ and thus has reduced radiance reflectance compared to the nearby 779.5 nm channel.

While the details will be different for an exoplanet, it is worth noting that, in both ultraviolet and visible channels, the Earth looks brighter in the southern hemisphere (SH) summer (February 8) when the southern oceans are facing more directly toward the EPIC camera, than in the northern hemisphere (NH) summer (August 8) when the northern landmass is more in the FOV. This is partly caused by the Earth being closer to the sun during the NH winter season, since its orbit has an eccentricity of ~0.017. This earth-sun distance change is good for a ~4% brightening of the SH summer compared to the NH summer. In addition, the larger Antarctica ice sheet contributes more reflected photons than the Arctic during their respective summers. Therefore, when we plot the time series for the entire dataset (Figure 4b and 4d), the peaks of the Earth's reflectance are shown during the SH summer (NH winter), in addition to the convective cloud maxima over the Western Pacific. Furthermore, a relatively small but clear radiance enhancement is also seen in the NH spring-summer period, likely due to the cloud enhancement caused by the South Asian monsoon.

The surface and cloud induced variabilities are shown in Figures 4g and 4h. The green color illustrates the surface-signature radiance, $I_s$, while the cloud-signature radiance, $I_c$, is shown with the blue color. Note that the pixel values of $I_s$ include both positive landmass and negative ocean contributions (Figure 3). Therefore, when averaging the full image $I_s$ into a single-point, it includes both land and ocean contributions, where the actual value depends on their relative area coverages. For Earth, the averaged $I_s$ values are mostly negative, since most of the Earth's surface is covered by oceans where the 780 nm surface reflectance is almost zero. The values of $I_c$ are all positive. Negative $I_s$ is an algebraic result of using $I_{388}$ to minimize cloud effects.

For surface variability, it can be seen that $I_s$ reaches its maximum on August 8 when both the African and Asian continents are in the FOV at ~10:00 UTC. On February 8, $I_s$ is maximized between 12:00 and 14:00 UTC when Africa and South America are both facing the EPIC camera (Figure 4g). For the entire time series (Figure 4h), $I_s$ peaks during the NH summer when most of the NH landmass is tilted toward the EPIC camera's FOV.

For cloud variability, a strong convective cloud system is the cause of the $I_c$ maximum enhancement at ~2:00 UTC on February 8, when the large convective clouds over the Western to Central Pacific are facing the EPIC camera. Similarly, $I_c$ peaks at ~ 3:00-4:00 UTC on August 8, when the Indian Monsoon region is in the EPIC's FOV. A secondary maximum occurs at 17:00 UTC on February 8 and 18:00 UTC on August 8, which is related to the convective clouds over the South America and Africa in the EPIC's FOV (Figure 4g). For the entire time series (Figure 4h), $I_c$ is enhanced during both the SH summer and NH summer, when strong convective cloud systems over the Western Pacific (during SH summer) and Asian Monsoon (during NH summer) result in strong enhancements of reflected radiances.

## 5. Fourier Analysis

Following the creation of a single-point source version of the EPIC dataset, we perform a Fourier analysis (Figure 5) to decompose EPIC's reflected radiance time series into components of different frequencies so as to explore its frequency composition. Similar Fourier analysis will apply to exoplanet observations, but with a different phase angle (~90°, see Section 6) than the Earth observations by EPIC (~180°).

First, a *Lomb-Scargle Periodogram* (Lomb, 1976; Scargle, 1982; Hans et al., 1999) is calculated for the time series of the channels of interest to account for the irregularities in temporal sampling. Although the EPIC instrument takes images at roughly equal time intervals each day, sometimes there are data gaps and the time interval changes over different seasons (Herman et al.

2017). More than 27 months of available EPIC data are used for this study, spanning more than 860 days, both daily and sub-daily frequencies were examined in high detail.

The results of the Fourier analysis illustrate the rich variety of information that can be obtained about planet Earth from time series of single-point measurements. Here, we choose three wavelength channels to focus our discussion: $I_{318}$, where signatures from clouds are strong (Figure 5a), $I_{388}$, where the cloud contribution dominates (Figure 5b) and $I_{780}$ nm, where land and clouds contribute almost equally (Figure 5c). We also show power spectra of $I_s$ (green) and $I_c$ (blue), where the signatures from surface and clouds are, respectively, the most significant (Figure 5d). The power spectra of other EPIC wavelength channels are shown in the Figure A2 in the Appendix.

The Fourier analysis technique decomposes the time series of EPIC single-point measurements into the frequencies or periodic signals that comprise it. The most significant feature is a large amplitude 1-day (24h) oscillation, shown as a power spectrum peak at the 24h period mark in Figure 5 and seen prominently in the Fourier results for all EPIC wavelength channels. This feature is almost certainly related to the Earth's 24-hour rotation period, since the main features of the Earth reappear every 24 hours.

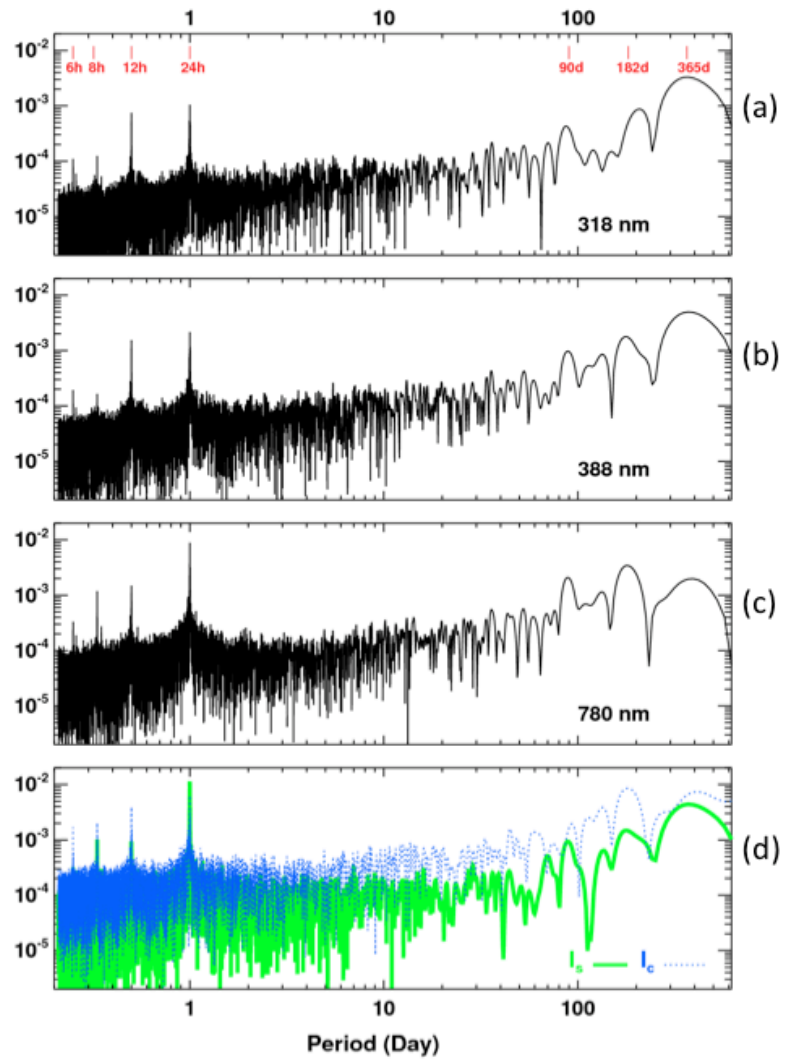

**Figure 5.** (a), (b) and (c) show Fourier series power spectra of DSCOVR EPIC L1B $I_{318}$, $I_{388}$ and $I_{780}$ radiances after averaging to the single-points; (d) shows the Fourier series power spectra of $I_s$ and $I_c$.

A few smaller amplitude features with shorter than 24-hour period are notable at 12-hour, 8-hour and 6-hour marks, which also appear at all wavelengths, but the 12-hour period has a larger amplitude in the ultraviolet channels (e.g. Figures 5a and 5b) and in $I_c$. These harmonics in the Fourier power spectra are likely related to periodic patterns of the Earth's surface and cloud systems, which repeat daily when the Earth rotates. This full phase light curve spectra are slightly different from the simulated spectra for a half-moon phase by Palle et al. (2008), where the peak at 12 hour can be higher than the peak at 24 hour and an autocorrelation method has been employed to find the rotation period of the planet.

The strong power spectrum at a period of 12-hours could be a combined signature from clouds, oceans and landmass patterns: (1) From DSCOVR's $L_1$ location, the Western Pacific is facing the EPIC camera at ~2:00-3:00 UTC and about 12 hours later at 14:00-15:00 UTC the Atlantic Ocean is in the FOV (Figure A1 in the Appendix); (2) Two major continents (Asia and North America) pass through the EPIC FOV in ~12-hour interval: China (Beijing) is 12 hours ahead of U.S.A (Washington, DC); (3) Australia is also 12 hours away from South America (Appendix Figure A1, right-panel). Thus, the 12-hour recurrence of cloud systems, ocean and large landmass views, and reflectance from the associated convective clouds in EPIC's FOV, all contribute to the 12-hour oscillation signature seen in the Fourier results. For example, all of the channels are known to receive a significant contribution from specular ocean reflections near the sub-satellite region. For $I_c$, oceanic convective cloud system could play a role.

The two smaller harmonics with 8-hour and 6-hour periods could also be related to the spatial variation of cloud system and the continents. For example, South America is 6 hours ahead of Southern Africa across the Atlantic, and Australia's west coast is 8 hours ahead of Africa (Cairo, Egypt), and South America and Africa are passing the EPIC FOV in 6-hour intervals, etc.

Longer than 24 hours, a few periodic signatures are seen in the Fourier power spectra. A maximum amplitude centered at the 365-day period, which is also seen at all wavelengths, is clearly related to the Earth orbiting the Sun, where the Earth orbit inclination angle relative to the Sun is ±23°, and the EPIC camera changes annually correlated with the hemispheric seasons (Figure A3 in the Appendix). Many natural events that have annual cycles, such as the Asian monsoon, tropical storms, dry and wet seasons in the Amazon, and re-occurrence of the Antarctica ice-cap in the EPIC FOV (e.g., Figure A4 in the Appendix), and others, contribute to the 365-day periodic signature. For $I_s$, this peak mainly comes from the asymmetry of land/ocean distribution between the SH and NH and the seasonal variation of the viewing angle due to the inclined Earth orbit.

Perhaps the most notable difference between the surface and cloud features is seen in the other two periodic harmonics with 180-day and 90-day periods. These are mainly due to the DSCOVR spacecraft's 6-month (180 days) orbit about $L_1$, which means that every 3 months (90 days) it is on the opposite east-west side of the Earth-Sun line (https://eosweb.larc.nasa.gov/project/dscovr/DSCOVR_overview_2016-06-29.pdf). In addition, the sub-solar position passes over the equator twice per year at the equinoxes, so the deep tropics have a 180 days (6 month) seasonal cycle. The 90-day cycle is seen more clearly in the $I_s$ and $I_{780}$ radiances, which might contain a component of the changes in the growth cycle of crops. Most of the vegetable crops have a ~90-day growth cycle (https://extension.umd.edu/hgic/plants/vegetable-crops), changing the greenness of the planet.

Finally, there seem to be no clear indications of the Moon's signature in the Fourier results. This is because the Earth and Moon must be in EPIC's FOV to be seen, which does not happen often. The Earth orbits the Sun in the ecliptic plane. The Moon's orbit around the Earth is inclined to the ecliptic by 5°9′. From its location at the L1 point, EPIC is pointed directly at the Earth and has a FOV of 0.61°, while the Earth subtends an angle of 0.45° to 0.53° depending on DSCOVR's six-month orbit. Thus, the Earth takes up almost the entire FOV. That FOV is much narrower than the ±5° motion of the Moon relative to the ecliptic plane, which means the Moon is rarely along the EPIC line of sight. Only twice since the launch of DSCOVR has the Moon been near the ecliptic plane and in EPIC's FOV 4° to 15° away from the Earth-Sun line. The two dates were July 16, 2015 and July 5, 2016, when EPIC captured the far side of the Moon's image moving across the Earth's illuminated face. However, during these two days, only RGB (red, green, and blue) visible images were taken and released to the general public (https://www.nasa.gov/feature/goddard/2016/nasa-camera-shows-moon-crossing-face-of-earth-for-2nd-time-in-a-year), and thus the EPIC L1B retrieval data are not available. Since the EPIC team is no longer doing this, the 10-channel L1B data will be produced for the next moon in the FOV. Occasionally, when the Moon is in the ecliptic plane, but not exactly in the EPIC FOV, it can cast a shadow projecting onto the Earth causing a solar eclipse where the shadow is cast. An example of the Moon's shadows in a 780 nm EPIC image, taken during the famous total solar eclipse on August 21, 2017, is shown in Figure A5 of the Appendix. Although the Moon passes the node of the ecliptic plane between the Earth and Sun every 27 days, a solar eclipse does not happen every month since the Moon may not be on the Earth-Sun line. It is for this reason that the moon's signature can rarely be seen in the Fourier analysis. Nevertheless, in the images of distant exoplanets, light from exomoons could certainly appear in the FOV of a camera. The potential influence of moonlike companions in the exoplanet data has been examined in a number of recent studies, e.g. Robinson (2011), Rein et al. (2014), and Agol et al. (2015).

## 6. Simulating the Phase Angle Effect

Since DSCOVR EPIC always views the sunlit side of the Earth because the spacecraft goes around the $L_1$ point, the above analysis of using EPIC data as a proxy for exoplanets are for near full phase configuration, which correspond to configurations when exoplanets are very close to host stars, thus hard to observe using coronagraph or starshade (Cash 2006). Most of the time, exoplanet observations see the planet with partial illumination, not the full starlit planet. The light curves from exoplanet missions would typically have phase angle effects, which we now study. To test how the phase angle change could impact our analysis, we artificially shade the EPIC images to simulate the changing phase angle that might be seen when viewing an exoplanet.

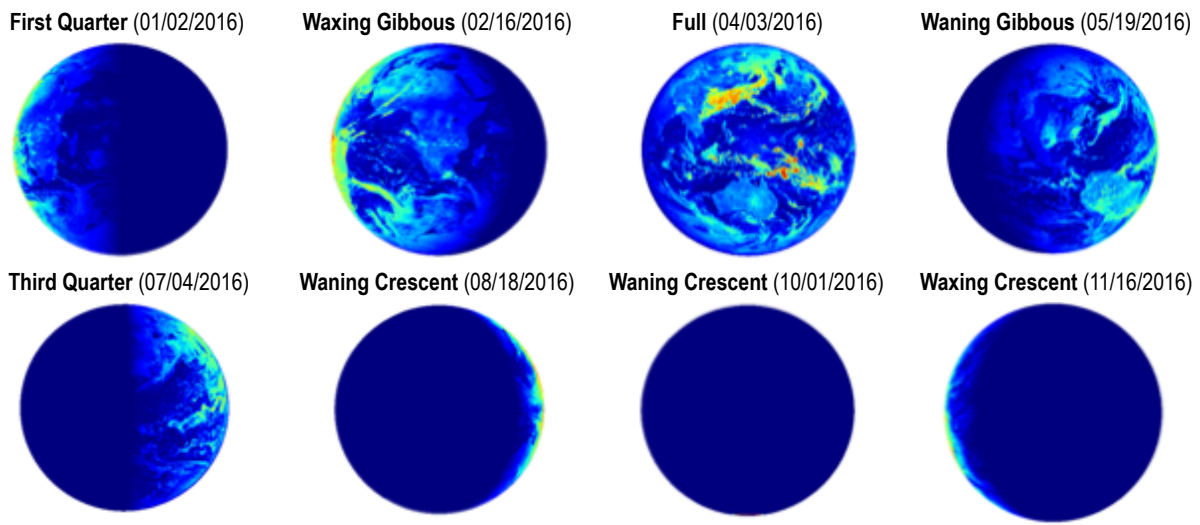

**Figure 6.** The simulated Earth image with edge-one phase angle changes.

First, we simulate viewing the orbit *edge-on*, which is the case for all the exoplanets detected by the transit method. We assume that the observed reflectance is independent of the incident angle, e.g. Lambertian and set the sun to be directly left of Earth on January 1 of each year. For simplicity, we assume that the Earth is spherical. To simulate phase angle effect, we first divide out the effect of the original illumination field, which fades away with the cosine of the angle from the center of view. At each phase angle, the star light comes from a direction perpendicular to the terminator plane, which separates day and night. For the day time side, the desired illumination factor at each pixel is then given by the ratio of the distance to the terminator plane to the radius of the sphere. After dividing out the original illumination, we multiply the illumination field calculated based on the terminator plane rotated according to the simulated phase angle for the date in a year of each image. Additionally, for each image, all the pixels representing the night side are set to zero. The duration of a complete phase angle change cycle is one full year (365 days), with the phases 45° apart as the Earth viewed as an exoplanet moves around the Sun (Figure 6).

The Fourier series power spectra of the point-light EPIC data with the edge-on phase angle simulated are shown in the Appendix Figure A6. Compared to the original Fourier power spectra, the amplitude of the 365-day peak is enhanced, signifying the period of the phase angle change as the Earth (proxy exoplanet) orbits the Sun (proxy star). The maxima at periods ≤ 24-hour are now mostly standing out in the red and near-infrared channels (> 551 nm) but somewhat weakened in the ultraviolet and visible ranges (318-551 nm). However, the main signals we see from the original full sunlit version are unchanged.

We also simulated another case, where the orbit is viewed face-on and the terminator boundary cuts directly through the center of the planet. The phase angle is again calculated from the date of each image, and then a line perpendicular to the position vector from the sun to Earth and crossing the center of Earth is drawn (Figure A7 in the Appendix). This is not a common case because so far most of the exoplanets were discovered by transit or radial velocity methods, which cannot detect exoplanets whose orbit is seen face-on. Also, the simple geometry dictates that orbital inclinations are usually closer to edge-on than face-on, for the same reason that half of Earth's area is within 30 degrees of the equator. Nevertheless, in this less likely case the Fourier series power spectra contains useful information (See Appendix Figure A8 and the associated discussion in the figure caption).

**7. Testing the minimum data collection rate for determining diurnal cycle**

For exoplanet observations, we expect the data collection rate to be much lower than what was obtained from the DSCOVR. An exoplanet's orbital period around the host star usually can be determined by measuring transits or starlight wobbles via variations in the light from the host star, or by planetary astrometry, when direct exoplanet images become available. However, detecting the rotation period is much more difficult because it is necessary to observe the light reflected from the exoplanet, which is a billion times fainter. Therefore, a single-point measurement requires a long integration time.

Here we seek to determine the minimum rate of data collection necessary to determine the rotation period or diurnal cycle of the Earth over a week of measurements. The EPIC instrument collects measurements every 6483 seconds, so data collection rates were tested with intervals that are integer multiples of 6483 seconds. The data were divided into 100 week-long segments. For each collection rate, we randomly create a sampling from every week-long segment of the time series of single point lights (specifically 780 nm) to create a total of 100 samples. Then, we calculate the Fourier power spectrum for each week-long segment and record whether or not the strongest peak in the power spectrum is at the frequency of the diurnal cycle range from 0.95 days$^{-1}$ to 1.05 days$^{-1}$. In this manner, for each collection rate, we estimate the probability of being able to discern the diurnal cycle by examining the frequency of the highest peak of the power spectrum.

**Table 2.** Data collection rate for determining the diurnal cycle of planet Earth

| Sampling Interval (hour) | Number of Samples Per-Day | Probability of Detecting Diurnal Cycle |
|---|---|---|
| 1.8008 | 13.3272 | 0.99 |
| 3.6017 | 6.6636 | 1.00 |
| 5.4025 | 4.4424 | 1.00 |
| 7.2033 | 3.3318 | 0.94 |
| 9.0042 | 2.6654 | 0.93 |
| 10.8050 | 2.2212 | 0.80 |
| 12.6058 | 1.9039 | 0.60 |
| 14.4067 | 1.6659 | 0.65 |
| 16.2075 | 1.4808 | 0.56 |
| 18.0083 | 1.3327 | 0.51 |
| 19.8092 | 1.2116 | 0.47 |
| 21.6100 | 1.1106 | 0.23 |
| 23.4108 | 1.0252 | 0.32 |
| 25.2117 | 0.9519 | 0.28 |
| 27.0125 | 0.8885 | 0.25 |
| 28.8133 | 0.8329 | 0.12 |
| 30.6142 | 0.7840 | 0.24 |
| 32.4150 | 0.7404 | 0.10 |
| 34.2158 | 0.7014 | 0.07 |
| 36.0167 | 0.6664 | 0.07 |

The testing results shown in Table 2 illustrate how often one needs to collect images in order to reliably detect the diurnal cycle, where the first column is the measurement frequency expressed in hours between measurements, the second column is the same frequency expressed in measurements per day, and the third column is the probability of having a power spectrum that contains the diurnal cycle as the strongest peak. Since these probabilities are derived from only 100 samples, the probability has an uncertainty of about a few percent on average according to a binomial distribution. This is the reason why at the shortest sampling of 1.8 hours we have 1 missing case, while for cases of 3.6 hours and 5.4 hours, there is no missed detection. In this testing for Earth, one needs to take single point light measurements at least every 9 hours in order to detect the Earth's rotation with a success rate of more than 90%. By the Nyquist theorem, if we sample the time series with an interval shorter than half of the period of a signal, we should be able to resolve the signal. True random noise due to photon shot noise and detector noise are very small compared with the signal for the DSCOVR data, so the reason we are not close to 100% probability to detect the diurnal cycle at sampling intervals of 10.8 hours and 9 hours, is most likely the variations of clouds.

Realistic exoplanet light curves are expected to have a much higher noise level, so it would be interesting to study how the noise affects this detection. Also it should be noted that for exoplanet observation, the sampling interval can be impacted by the integration time. For example, if the integration time is 6 hours, then it means that the data are 6-hour averages, rather a snapshot taken every 6 hours. This might impact one's ability to detect rotational variability. We defer these topics to a future study given there will be more data for better statistics.

## 8. Summary and Discussion

In this study, more than two years of Earth images taken from the $L_1$ point by the DSCOVR spacecraft were used for a novel experiment, in which the Earth served as a proxy exoplanet. We separated cloud and surface (land and ocean) signatures in the DSCOVR EPIC images using a simple linear combination of two wavelength channels to remove cloud contribution. We averaged the Earth resolved images to single-point "measurements" and used Fourier analysis to obtain information about the Earth's rotation, its orbit around the Sun, and possible periodic variations due to weather (clouds) patterns, surface type (ocean, land, vegetation), and season. These results were validated by comparing them with known data about our planet. We further simulated phase angle changes, and the minimum data collection rate needed to determine the rotation period of an exoplanet. This method of using the time evolution of a multi-wavelength, reflected single-point light source, can be deployed for retrieving a range of intrinsic properties of a true exoplanet around a distant star.

We note that caveats exist when applying our methodology to exoplanet applications. The most obvious one is that the data we used is specific to Earth's atmosphere and surface. It is likely that an exoplanet could have a completely different atmospheric composition, mass, and dynamics, as well as surface features. Nonetheless, if there is an atmosphere, there will be Rayleigh scattering, and probably clouds, even if not water clouds. While the details of our method based on DSCOVR may be mainly applicable to a water-based planet with clouds of moderate temperature (not frozen), we should see analogous effects from an ammonia-methane based planet similar to Uranus, or the hydrocarbon atmosphere and surface observed during Huygens's January 2005 landing on Saturn's moon Titan. Our analysis also assumes that the atmosphere is partially transparent in the visible range. If there is no Earth-like plant life, then the red-edge brightening feature at > 700 nm would be absent. But, water will still be dark in the near-infrared compared to rocky soil without vegetation. While the method should be generally applicable, care must be taken in interpreting the results from a Fourier analysis when applied to planets with different atmospheric and surface properties. In addition, we use averaging rather than integration to produce the "single-point" light sources from EPIC data. This is because the number of pixels covered by the Earth in each EPIC image is not always the same due to the slight eccentricity of the Earth's orbit and the 6-month period oscillation in DSCOVR's orbital radius. Integrating the observed signal is closer to an exoplanet observation, but the difference is only one of a scale factor.

Despite these caveats, any periodically varying features on the surface or the atmosphere of an exoplanet, such as a lasting "hurricane" similar to the Great Red Spot on Jupiter, would be certainly detectable using Fourier analysis should a long exoplanet time series of direct imaging data become available. Using the time evolution of multi-wavelength reflected light to extract various intrinsic properties of a planet would therefore open doors for future direct imaging missions to extract fundamental properties of exoplanets that could be key to modeling of their weather and climate. These could include the planetary rotation, seasonal changes of cloud, land and ocean features, exomoon information, and orbital characteristics. The parameter extraction and Fourier analysis techniques we are exploring may be useful to inform future mission design on the required cadence and image bands needed to constrain these important planetary parameters.

The TPF mission was canceled by NASA in 2012; The ESA's Darwin mission study was ended in 2007, and no further activity has been planned since then. However, the James Webb Space Telescope (JWST), is scheduled for launch in 2019. It will be located at the Earth-Sun $2^{nd}$ Lagrange point ($L_2$) — another quasi-stable orbital location 1.5 million km from Earth, just behind its shadow. JWST will be capable of pursuing many astronomical research topics and studying exoplanets is one of its primary areas of investigation. Another planned mission is the Wide-Field Infrared Survey Telescope (WFIRST), which will sharpen our ability to capture actual images of distant planets using an internal coronagraph instrument to selectively block and process incoming starlight to reveal the planets hidden in the glare of stars. In addition, a unique device outside the telescope called the star-shade (Cash 2006) has been developed at the Jet Propulsion Laboratory, California Institute of Technology. This device would be deployed in deep space as a giant structure, a few tens of thousands of kilometers away from a space telescope and pointing towards it, to block the glare of starlight. This will allow images of exoplanets around the target star to be captured. In addition, the NASA-sponsored missions such as the Habitable Exoplanet Imaging Mission (https://www.jpl.nasa.gov/habex/) and the Large Ultraviolet, Optical, and Infrared Surveyor (https://asd.gsfc.nasa.gov/luvoir/) have been recommended by the Astrophysics Decadal Survey.

In the photograph of planet Earth taken by the Voyager 1 spacecraft from 6 billion kilometers away on February 14, 1990, the Earth appears as a single-point light, a Pale-Blue Dot (http://www.planetary.org/explore/space-topics/earth/pale-blue-dot.html). In the coming decades, as space telescopes become larger and more sophisticated, direct imaging of another Earth will become possible. The results from DSCOVR will serve as a guide experiment to provide an effective baseline, from which tools can be developed and applied to a variety of exoplanet imaging data, extracting information about faraway worlds with continents, clouds and oceans.

**Acknowledgement:** This work was support by the Exoplanet Study Initiative (ESI) at the Jet Propulsion Laboratory (JPL), California Institute of Technology (Caltech), under contract with NASA. We acknowledge the DSCOVR project science team for support. We thank Nicolas Cowan of McGill University for detailed and constructive comments. Anthony Davis and Gerard van Harten of JPL, Stuart Bartlett of Caltech, Sara Seager of Massachusetts Institute of Technology, and Adam Showman of University of Arizona also provided useful comments on the data calibration, analysis methodology, and exoplanet imaging techniques.

**Data and code availability:** The DSCOVR data used for this study can be downloaded at https://eosweb.larc.nasa.gov/project/dscovr/dscovr_epic_l1a_2. The computer code used during the study are available on request from the authors. Please contact the corresponding author at Jonathan.H.Jiang@jpl.nasa.gov.

# Appendix

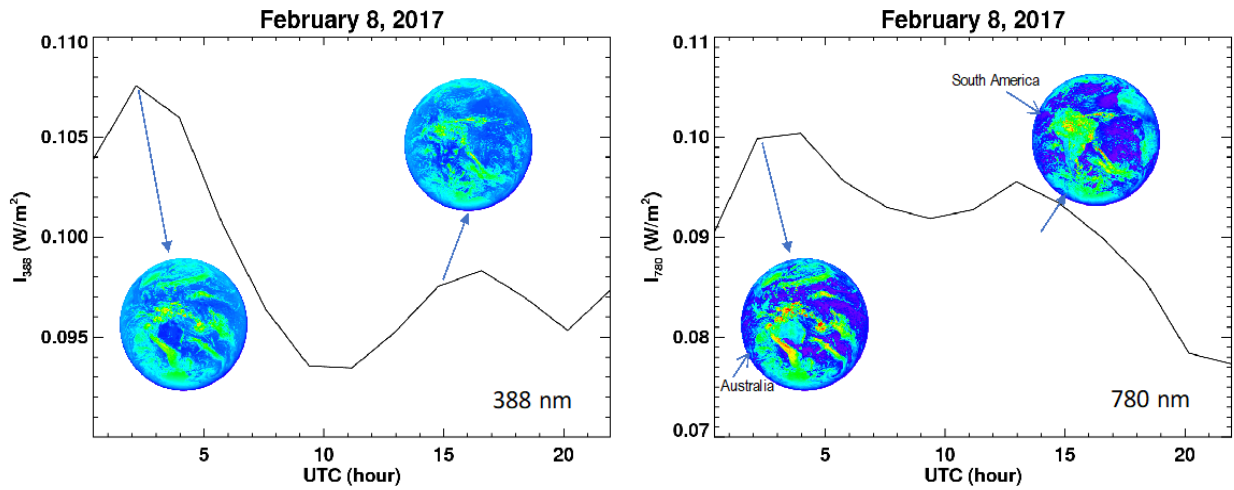

**Figure A1:** Variation of single-point averaged Earth reflectance on February 8, 2017 for I388 (left-panel) and I788 (right-panel) radiances, respectively. Note that the image of Australia (and the Pacific Ocean) is 12 hours apart from the image of South America (and the Atlantic Ocean).

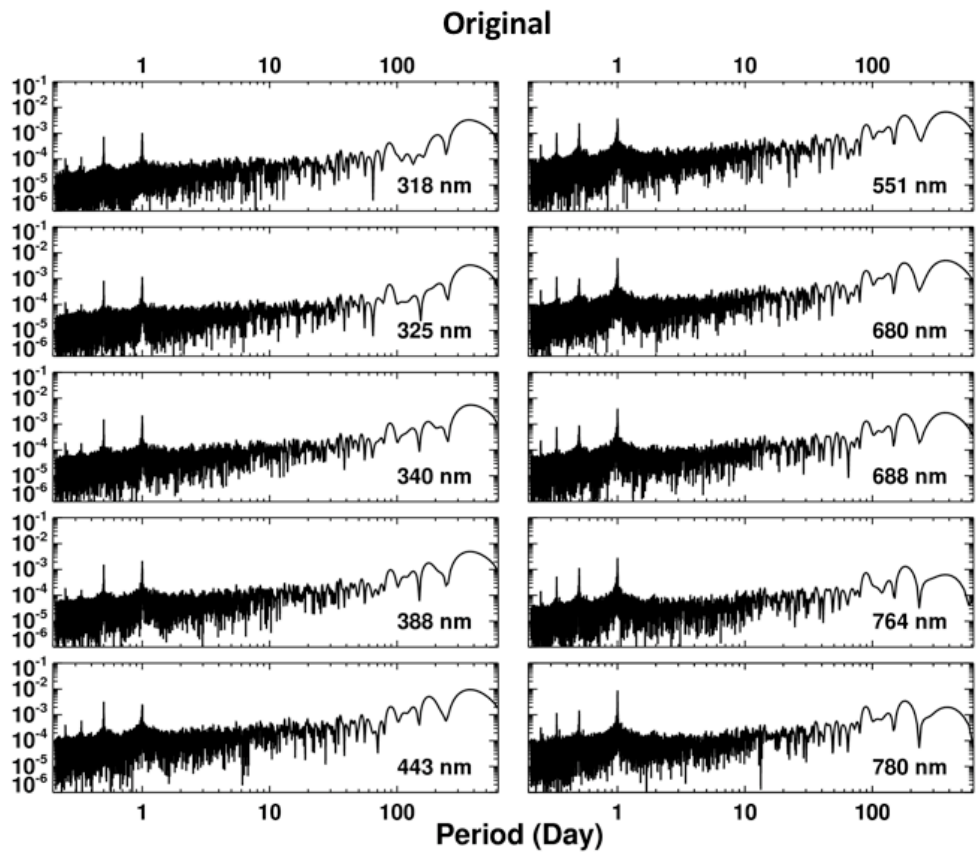

**Figure A2.** Fourier series power spectra of DSCOVR EPIC L1B radiances at the 10 EPIC wavelength channels after averaging them to the single-points.

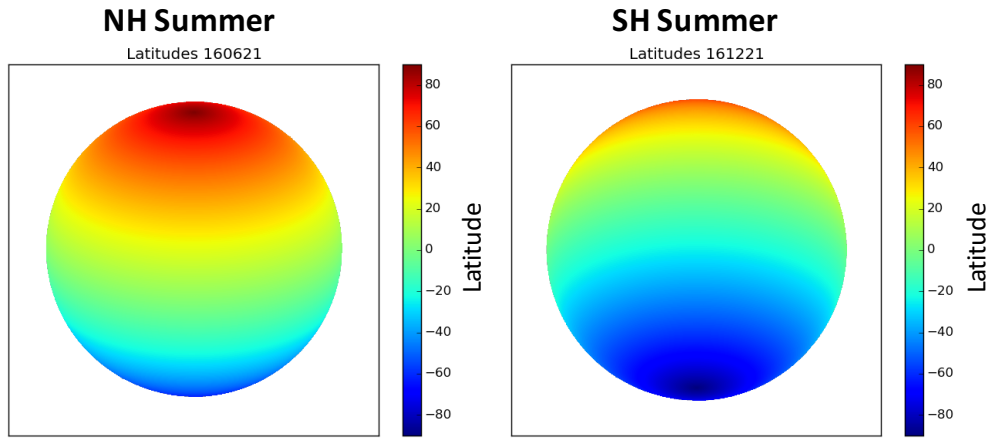

**Figure A3:** The geolocation reference of the Earth as seen from the DSCOVR's L1 location on June 21 (left-panel) and December 21 (right-panel), 2016. The Earth's orbital inclination angle changes annually relative to the Sun and the EPIC camera.

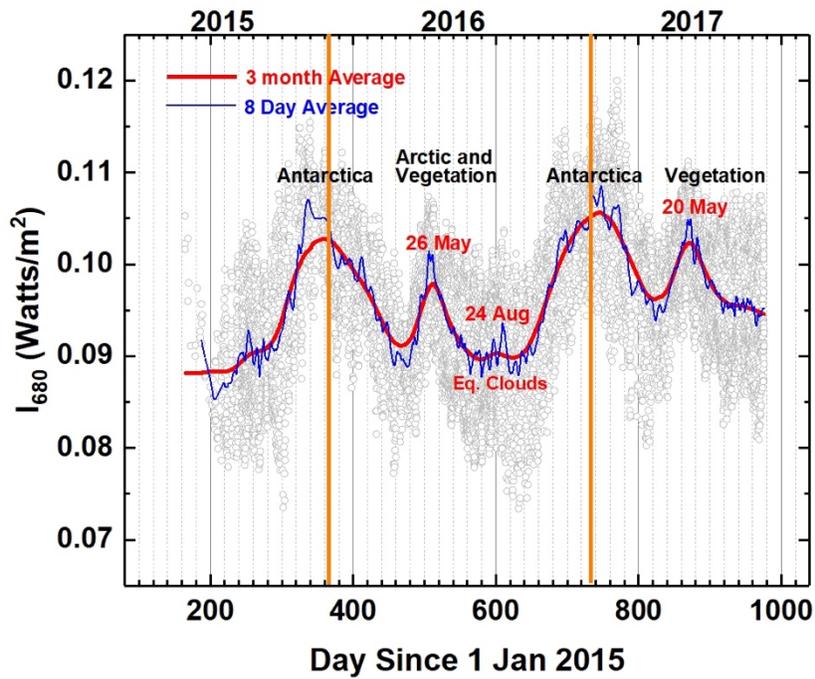

**Figure A4:** Time series of EPIC L1B 680 nm single-point radiances. The annual recurrence of typical surface features such as the Antarctic ice-cap is evident.

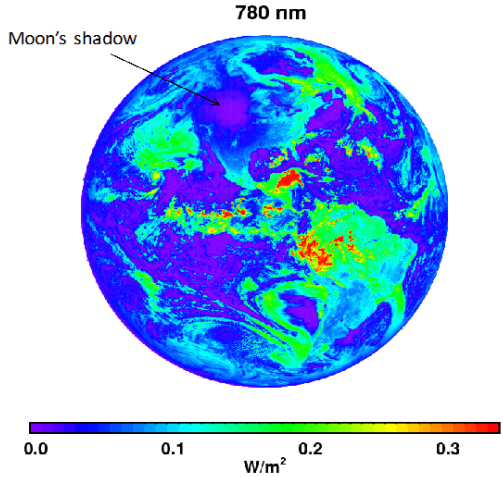

**Figure A5:** The Moon's shadow seen as a reduced radiance spot over the central U.S. on the luminous surface of the Earth in each of EPIC images taken on August 21, 2017, 17:42:36 UTC.

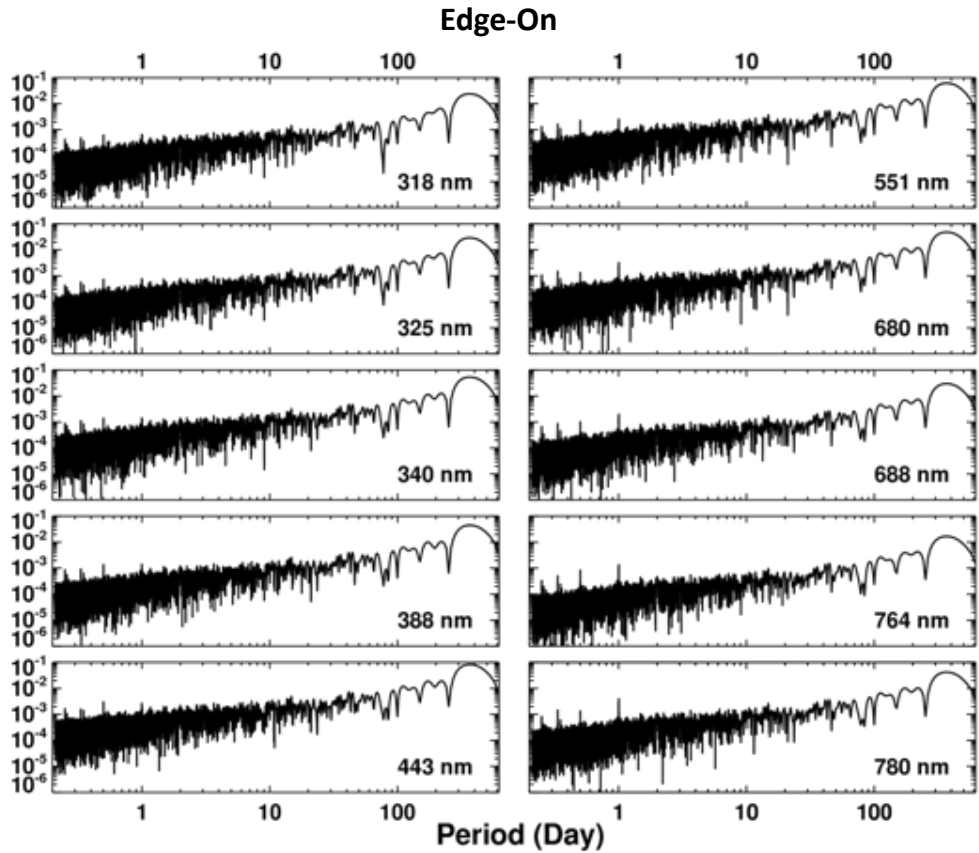

**Figure A6:** Fourier series power spectra of DSCOVR EPIC L1B radiances with *edge-on* phase angle change simulated at the 10 EPIC wavelength channels after averaging them to single-points. Compared to the original Fourier power spectra, the amplitude of the 365-day peak is enhanced, signifying the period of the phase angle change as the Earth (exoplanet) orbits the Sun (star). The maxima at periods less than 24 hours are now mostly standing out in the red and near-infrared channels (> 551 nm) but somewhat weakened in the ultraviolet and visible (318-551 nm). However, the main signals we see from the original full sunlit remain unchanged.

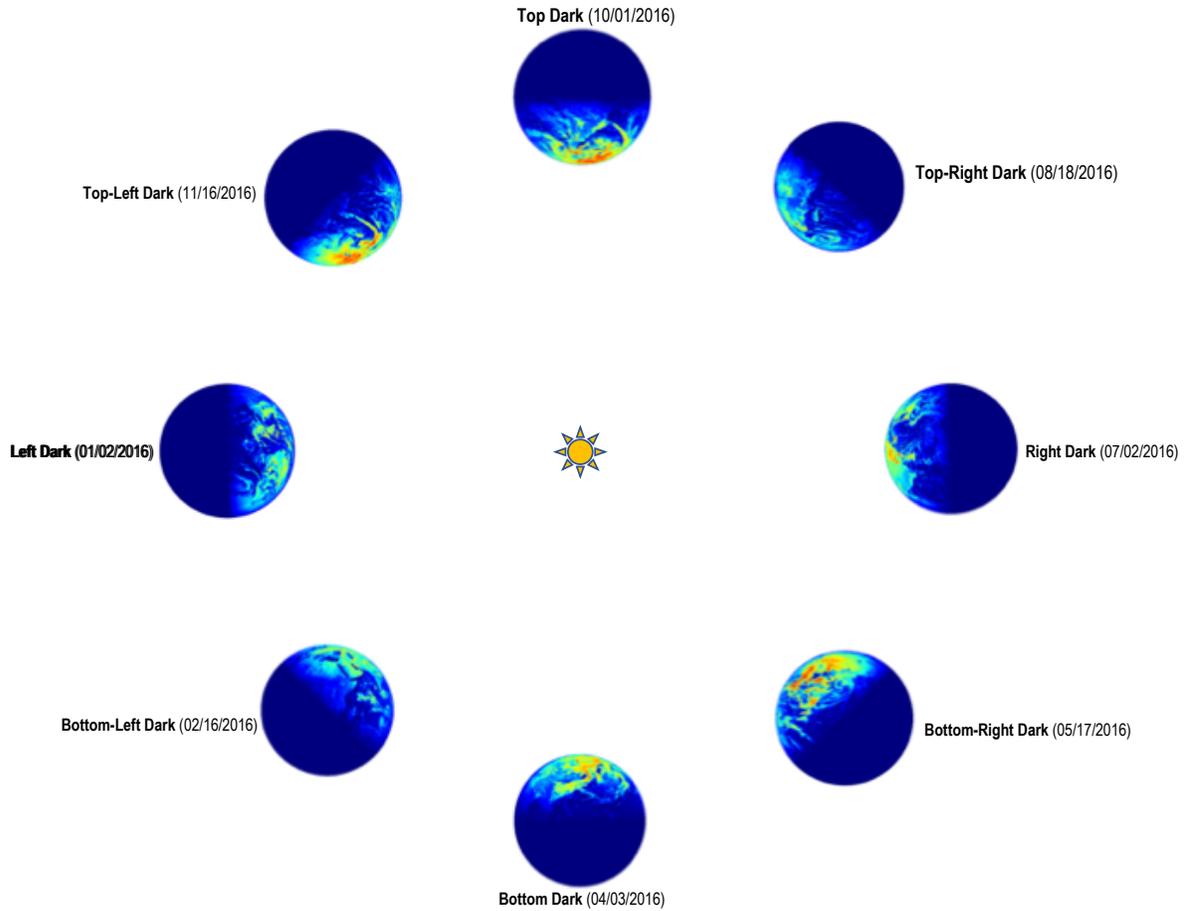

**Figure A7:** The simulated Earth image with face-on phase angle changes. This face-on case, however, is unlikely to be common since most of the exoplanets were discovered by transit or radial velocity methods, which cannot detect an exoplanet whose orbit is seen face-on.

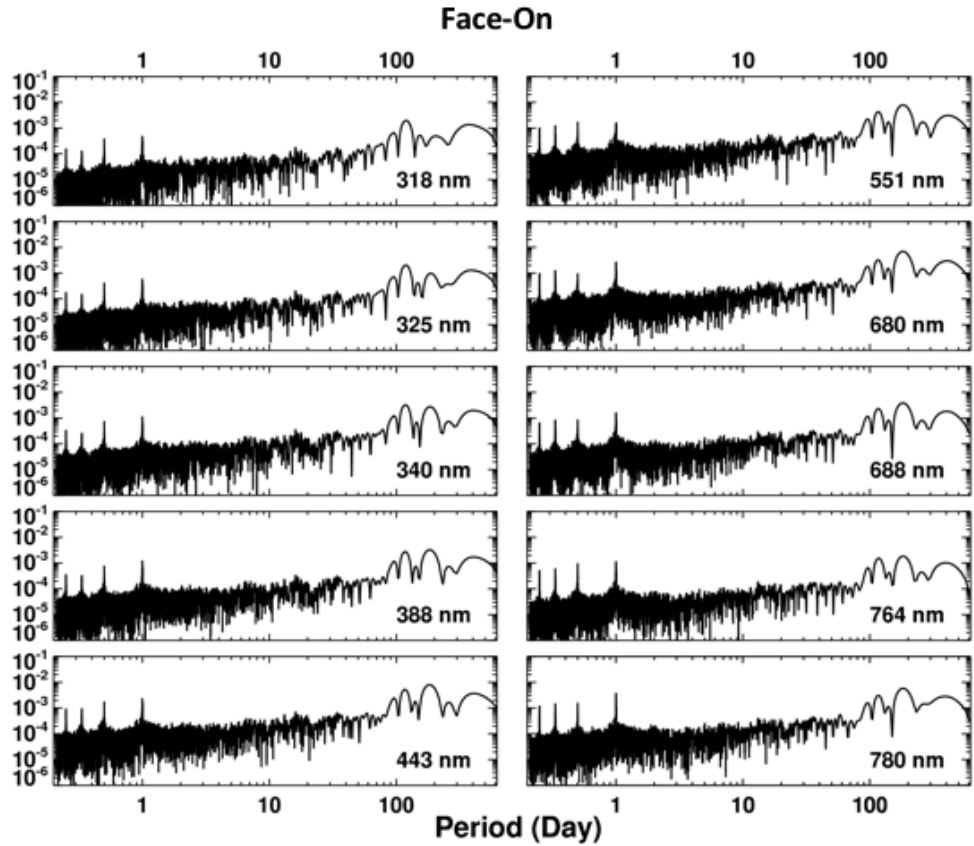

**Figure A8:** Fourier series power spectra of DSCOVR EPIC L1B radiances with face-on phase-angle simulated at the 10 EPIC wavelength channels after averaging them to the single-points. Compared to the original Fourier power spectra, the maxima at periods less than 24 hours are now mostly matching in all wavelength channels. The main signals we see from the original full sunlit remain unchanged.